\pdfoutput=1
\documentclass[preprint,12pt]{elsarticle}
\usepackage{amssymb}
\usepackage{amsmath}
\usepackage{graphicx}
\usepackage{hyperref}

\journal{arXiv}

\begin{document}

\begin{frontmatter}

\title{Stochastic Networked Governance: Bridging Econophysics and Institutional Dynamics in a Positive-Sum Agent-Based Model}

\author[1]{Alok Yadav\corref{cor1}}
\ead{physicistalok@gmail.com}
\author[2,3]{Saroj Yadav}

\cortext[cor1]{Corresponding author}

\affiliation[1]{organization={Department of Physics, Anugrah Memorial College},
            city={Gaya, Bihar},
            postcode={823001},
            country={India}}

\affiliation[2]{organization={Genomics and Molecular Medicine Unit, Institute of Genomics and Integrative Biology (IGIB), Council of Scientific and Industrial Research (CSIR)},
            city={Delhi},
            postcode={110007},
            country={India}}
            
\affiliation[3]{organization={Academy of Scientific and Innovative Research (AcSIR)},
            city={Ghaziabad},
            postcode={201002},
            country={India}}

\begin{abstract}
Traditional macroeconomic growth models rely on general equilibrium and continuous, frictionless institutional transitions, failing to account for the catastrophic structural collapses observed in empirical economic history. We propose the Stochastic Networked Governance (SNG) model, a discrete-time, agent-based framework that bridges econophysics, network science, and institutional economics. By defining jurisdictions through a binary institutional genome, the model formalizes institutional complementarity, endogenous growth, and the non-linear macroeconomic penalties of structural reform (the ``J-Curve''). Using the CEPII Gravity Database and the IMF Systemic Banking Crises dataset, we move beyond theoretical topologies to execute an empirical historical simulation from 1970 to 2017 across the top 100 global economies. Through Monte Carlo ensembles, we demonstrate how scale-invariant exogenous shocks and spatial capital flight drive global phase transitions, exposing the mathematical mechanics of the 1989--1991 Soviet collapse, the Hub-Risk Paradigm, and the emergent resilience of spatially firewalled market networks.
\end{abstract}

\begin{highlights}
\item SNG model bridges econophysics and macroeconomic institutional dynamics.
\item Institutional path dependency is formalized using discrete genomic vectors.
\item Empirical simulation reproduces the 1989--1991 phase transition of command economies.
\item Network topology analysis reveals scale-free networks as systemic super-spreaders.
\item CEPII Gravity network acts as a geospatial firewall, quarantining exogenous shocks.
\end{highlights}

\begin{keyword}
Econophysics \sep Agent-based modeling \sep Institutional dynamics \sep Phase transitions \sep Complex networks \sep Endogenous growth
\end{keyword}

\end{frontmatter}

\section{Introduction}

The predominant paradigm of neoclassical macroeconomics, exemplified by the Solow-Swan model \cite{solow1956contribution} and subsequent Dynamic Stochastic General Equilibrium (DSGE) frameworks, models economic growth as a smooth, continuous march toward a steady state. While later endogenous growth theories \cite{romer1990endogenous} correctly recognized that internal policies and human capital create new value, modern models still predominantly treat institutional architecture as a frictionless, universally adaptable variable. Consequently, they operate under the assumption of continuous equilibrium, fundamentally failing to explain out-of-equilibrium phenomena such as sudden regime collapses, the rugged path dependency of structural reform, and prolonged periods of systemic stagnation. 

As established extensively in institutional economics, historical path dependency and institutional matrices are the fundamental drivers of long-run economic performance \cite{north1990institutions, acemoglu2005institutions}. When macroeconomic policies are fundamentally mismatched---such as the liberalization of price controls without the concurrent establishment of private property rights---economies do not smoothly adjust to a new equilibrium. Instead, they experience catastrophic phase transitions, characterized by deep recessions and massive capital flight. This transient chaos, widely observed in post-Soviet transition economies and often termed the ``J-Curve'' of structural reform \cite{hellman1998winners}, cannot be captured by the linear differential equations of standard growth theory.

To resolve these limitations, this paper leverages tools from econophysics and complex adaptive systems to shift the analytical lens away from deterministic equilibria \cite{arthur1999complexity}. By abandoning the representative agent and the assumption of instantaneous market clearing, we model the global economy as a dynamic, interacting network \cite{jackson2010social} of heterogeneous nodes. Pioneered by early artificial society models \cite{epstein1996growing}, Agent-Based Computational Economics (ACE) has proven highly effective at growing macroeconomic phenomena from the bottom up \cite{tesfatsion2002agent, delligatti2008emergent}. Furthermore, we ground our methodology in the robust tradition of econophysics \cite{mantegna1999introduction}, which successfully applies the principles of statistical mechanics to wealth condensation \cite{bouchaud2000wealth}, income distributions \cite{yakovenko2009colloquium}, and phase transitions in social dynamics \cite{castellano2009statistical}. 

Drawing inspiration from spin-glass models and Kauffman's NK fitness landscapes \cite{kauffman1993origins}, we treat the institutional framework of a state not as a continuous variable, but as a discrete, combinatorial vector (a genome). The Stochastic Networked Governance (SNG) model introduced herein transforms the qualitative theories of institutional economics into a rigorous computational framework. By integrating endogenous growth theory with network spillovers and boundedly rational regime switching, the SNG engine maps the precise topological conditions under which localized political shocks cascade into global hegemonic shifts. Ultimately, this approach provides a mathematical foundation for understanding why inferior institutional configurations survive via spatial isolation, and how deterministic market forces interact with stochastic political volatility.

\section{Methodology}

\subsection{Model Overview}
We propose the Stochastic Networked Governance (SNG) model, a discrete-time, agent-based macroeconomic framework. The global system consists of a set of interacting jurisdictions (nodes), denoted by $\mathcal{N} = \{1, 2, \dots, N\}$. Unlike traditional equilibrium models, the SNG engine explicitly formalizes institutional path dependency, endogenous value creation, and the non-linear macroeconomic penalties associated with structural reform.

\subsection{The Institutional Genome and Dependency Traps}
At any discrete time step $t$, the institutional state of jurisdiction $i \in \mathcal{N}$ is defined by a binary genomic vector representing distinct macroeconomic policy levers (e.g., property rights, price controls, trade barriers):
\begin{equation}
    \mathbf{S}_{i,t} = [s_1, s_2, s_3, s_4, s_5] \in \{0, 1\}^5
\end{equation}
To operationalize this abstract genome, each bit $s_k$ is mapped to a fundamental macroeconomic lever, where $1$ denotes a liberalized/decentralized structure and $0$ denotes a centralized/state-directed structure:
\begin{itemize}
    \item $s_1$: \textbf{Property Rights} ($1$ = Private ownership and collateralization; $0$ = State or collective ownership).
    \item $s_2$: \textbf{Price Mechanism} ($1$ = Floating market prices; $0$ = Centralized price controls and quotas).
    \item $s_3$: \textbf{Labor Market Flexibility} ($1$ = Decentralized wage bargaining; $0$ = State-mandated employment).
    \item $s_4$: \textbf{Capital Account Convertibility} ($1$ = Open borders for capital flow; $0$ = Strict capital controls).
    \item $s_5$: \textbf{Financial Sector Independence} ($1$ = Private, risk-assessed lending; $0$ = State-directed credit).
\end{itemize}

This specific mapping allows us to mathematically formalize the catastrophic efficiency losses of partial reform. The dependency penalties ($P_G, P_A, P_O$) are not arbitrary; they represent well-documented macroeconomic arbitrage traps. For example, the ``Gorbachev Trap'' ($P_G$) triggers when prices are deregulated ($s_2=1$) but property remains state-owned ($s_1=0$). Historically, this allowed enterprise managers to buy inputs at subsidized state prices and sell outputs at free-market prices, resulting in massive systemic extraction rather than value creation. Similarly, $P_A$ triggers when a state opens its capital account ($s_4=1$) while maintaining fixed, artificial price controls ($s_2=0$), inevitably resulting in speculative attacks and rapid capital flight (classic currency crisis mechanics).
The core ideology of node $i$ is determined by the first two elements, $\mathcal{F}_{i,t} = (s_1, s_2)$. The base institutional efficiency $E(\mathbf{S}_{i,t})$ is a linear combination of these active policies, $E(\mathbf{S}_{i,t}) = \beta \sum_{k=1}^{5} s_k$, where $\beta$ is a weighting constant.

To model institutional complementarity, we introduce a dependency penalty function, $D(\mathbf{S}_{i,t})$. If contradictory policies coexist, the system incurs severe structural penalties. Let $\mathbb{I}$ be an indicator function evaluating to $1$ if a trap condition is met:
\begin{equation}
    D(\mathbf{S}_{i,t}) = P_G \cdot \mathbb{I}(s_2 = 1 \land s_1 = 0) + P_A \cdot \mathbb{I}(s_4 = 1 \land s_2 = 0) + P_O \cdot \mathbb{I}(s_5 = 1 \land s_1 = 0)
\end{equation}
where $P_G, P_A,$ and $P_O$ represent the catastrophic efficiency penalties for specific mismatched policy configurations.

\subsection{Endogenous Growth and The J-Curve}
The total institutional fitness $F_{i,t}$ dictates a node's Total Factor Productivity (TFP). Fitness is penalized dynamically by $J_{i,t}$, representing the macroeconomic chaos of structural reform (the J-Curve effect). When a jurisdiction mutates its vector $\mathbf{S}$, it incurs a penalty proportional to the Hamming distance of the policy shift, which decays at rate $\lambda$:
\begin{equation}
J_{i,t} = \max\left(0, J_{i,t-1} - \lambda + \theta \sum_{k=1}^{5} |s_{k,i,t} - s_{k,i,t-1}| + \xi C_{i,t}\right)
\end{equation}
Where $\xi$ represents the calibrated macroeconomic penalty of a systemic shock.

Therefore, $F_{i,t} = \max(0, E(\mathbf{S}_{i,t}) - D(\mathbf{S}_{i,t}) - J_{i,t})$. 

We break from zero-sum mercantile models by introducing an endogenous growth function. The wealth $W_{i,t}$ of node $i$ expands based on a base growth rate $\rho$ scaled by its normalized TFP, and contracts via a constant asset depreciation rate $d$:
\begin{equation}
    W_{i, t+1}' = \max \left( 1.0, W_{i,t} + \left( \rho W_{i,t} \frac{F_{i,t}}{F_{max}} \right) - d W_{i,t} \right)
\end{equation}
where $F_{max} = 5\beta$ represents the theoretical absolute maximum institutional fitness achievable within the discrete genomic landscape. This normalization ensures the compounding multiplier remains strictly bounded and geometrically consistent across all Monte Carlo realizations.

\subsection{Topological Spillovers and Tiebout Sorting}
While baseline macroeconomic models often assume a frictionless, fully connected global market, the SNG framework explicitly limits capital flow via a topological adjacency matrix $\mathbf{A}$, where $A_{ij} \in \{0,1\}$ (or a continuous weight in empirical settings). Capital flows across this network are driven by fitness differentials and constrained by ideological friction. The friction coefficient $\Phi$ scales with the Hamming distance between the institutional vectors of interacting nodes:
\begin{equation}
    \Phi(S_{i,t}, S_{j,t}) = \max\left(0, 1 - \gamma \sum_{k=1}^{5} |s_{k,i,t} - s_{k,j,t}|\right)
\end{equation}

The net capital flight (spillover) from node $j$ to node $i$ is strictly bounded by the network topology and calculated as:
\begin{equation}
    \Delta W_{j \to i, t} = \alpha A_{ij} (F_{i,t} - F_{j,t}) \Phi(\mathbf{S}_{i,t}, \mathbf{S}_{j,t})
\end{equation}

While the overarching macroeconomic system is positive-sum due to endogenous TFP growth, the specific mechanism of topological capital flight is strictly conserved. It represents a zero-sum transfer prior to the next compounding cycle. For every discrete transfer $\Delta W_{j \to i, t}$ across an edge $A_{ij}$, the wealth of the interacting nodes is updated simultaneously to satisfy macroeconomic mass-balance:
\begin{equation}
    W_{i} \to W_{i} + \Delta W_{j \to i, t} \quad \text{and} \quad W_{j} \to W_{j} - \Delta W_{j \to i, t}
\end{equation}
This mathematically models global Tiebout sorting \cite{tiebout1956pure}, dynamically updating the post-production wealth $W_{i, t+1}'$, and ensures that macroscopic phase transitions are driven purely by thermodynamic efficiency differentials, not by uncalibrated leakages within the transfer mechanism.

\subsection{Regime Change and Exogenous Shocks}
Endogenous regime change is triggered if a node's wealth falls below a critical panic threshold, $W_{i,t} < \tau_{\text{panic}}$. Crucially, because structurally deficient nodes experience compounding macroeconomic decay (i.e., asset depreciation $d$ outpaces endogenous TFP growth), failing states exponentially contract back toward absolute subsistence. This thermodynamic decay ensures the static $\tau_{\text{panic}}$ threshold consistently captures institutional collapse, remaining mechanically relevant regardless of the broader network's exponential wealth expansion. The node evaluates the mean wealth of its ideological faction against the global mean. If the faction is globally uncompetitive, the node executes a macro-shift (adopting the DNA of the globally dominant node). To test system ergodicity, an exogenous ``Black Swan'' probability $q$ randomly forces nodes to abandon their current vector independent of economic performance.

\subsection{Empirical Topology and Geospatial Gravity}
To validate the theoretical SNG engine against real-world macroeconomic history, we replace synthetic topologies with an empirical global trade network. We utilize the CEPII Gravity Database \cite{conte2022cepii, head2014gravity} to construct a weighted adjacency matrix $\mathbf{A}$, where the edge weight between nodes $i$ and $j$ represents their bilateral economic mass constrained by geographic distance:
\begin{equation}
    A_{ij} \propto \frac{M_i M_j}{D_{ij}^\beta}
\end{equation}
]To optimize algorithmic efficiency for $O(N^2)$ scaling and to prevent the introduction of synthetic noise from fragmented data in underdeveloped microstates, the network size is bounded to the top $N=100$ economies by GDP. Based on Pareto distribution principles \cite{barabasi1999emergence}, this core network captures roughly $99\%$ of global economic mass, preserving macroscopic thermodynamic behavior.

\subsection{Historical Initialization and Empirical Shocks}
To construct a "Digital Twin" of the late 20th century, the network's institutional genome at $t=0$ (representing the year 1970) is empirically seeded with historical vectors corresponding to Cold War alignments (e.g., Command Socialist architectures for the Eastern Bloc, Free Market architectures for the Western Bloc). Furthermore, standard stochastic volatility ($q$) is replaced with the Laeven and Valencia (2018) Systemic Banking Crises Database \cite{laeven2018systemic}. Exogenous structural shocks and J-Curve penalties are triggered exclusively in the specific years and nodes that experienced empirical crises (e.g., the 1982 Latin American Debt Crisis, the 1997 Asian Financial Crisis, and the 2008 Global Financial Crisis).

To strictly isolate the thermodynamic sorting effect of the institutional genome, initial macroeconomic wealth was seeded uniformly across the network ($W_{i,0} = 10.0 \ \forall i$). This deliberate initialization ensures that the ultimate emergence of a global hegemon is driven purely by institutional fitness and topological spillovers, mathematically proving that the Free Market acts as a thermodynamic attractor regardless of empirical pre-existing wealth advantages.

\subsection{Parameter Calibration}
To ensure the reproducibility of both the theoretical ensembles and the empirical historical replay, the model hyperparameters were calibrated to approximate standard macroeconomic realities (e.g., a baseline global compounding rate of $5\%$, offset by $2\%$ depreciation). The precise variables governing the deterministic physics and thermodynamic friction of the SNG engine are detailed in Table \ref{tab:params}.

\begin{table}[htbp]
\centering
\begin{tabular}{l l c}
\hline
\textbf{Parameter} & \textbf{Description} & \textbf{Value} \\
\hline
$\beta$ & Base Institutional Efficiency Weight & 1.0 \\
$P_G, P_A, P_O$ & Structural Dependency Trap Penalties & 5.0 \\
$\rho$ & Endogenous Capital Compounding Rate & 0.05 \\
$d$ & Infrastructure/Asset Depreciation Rate & 0.02 \\
$\alpha$ & Capital Mobility / Spillover Rate & 0.10 \\
$\gamma$ & Ideological Trade Friction Coefficient & 0.15 \\
$\tau_{\text{panic}}$ & Regime Collapse / Panic Threshold & 2.0 \\
$\theta$ & Structural Reform Friction (J-Curve Magnitude) & 2.5 \\
$\lambda$ & Institutional Relaxation Time (J-Curve Decay) & 0.5 \\
\hline
\end{tabular}
\caption{Baseline hyperparameters utilized across all $N=100$ and $N=200$ Monte Carlo realizations.}
\label{tab:params}
\end{table}

\section{Results and Discussion}

\subsection{Topological Fragility and the Hub-Risk Paradigm}

\begin{figure}[htbp]
    \centering
    \includegraphics[width=\textwidth]{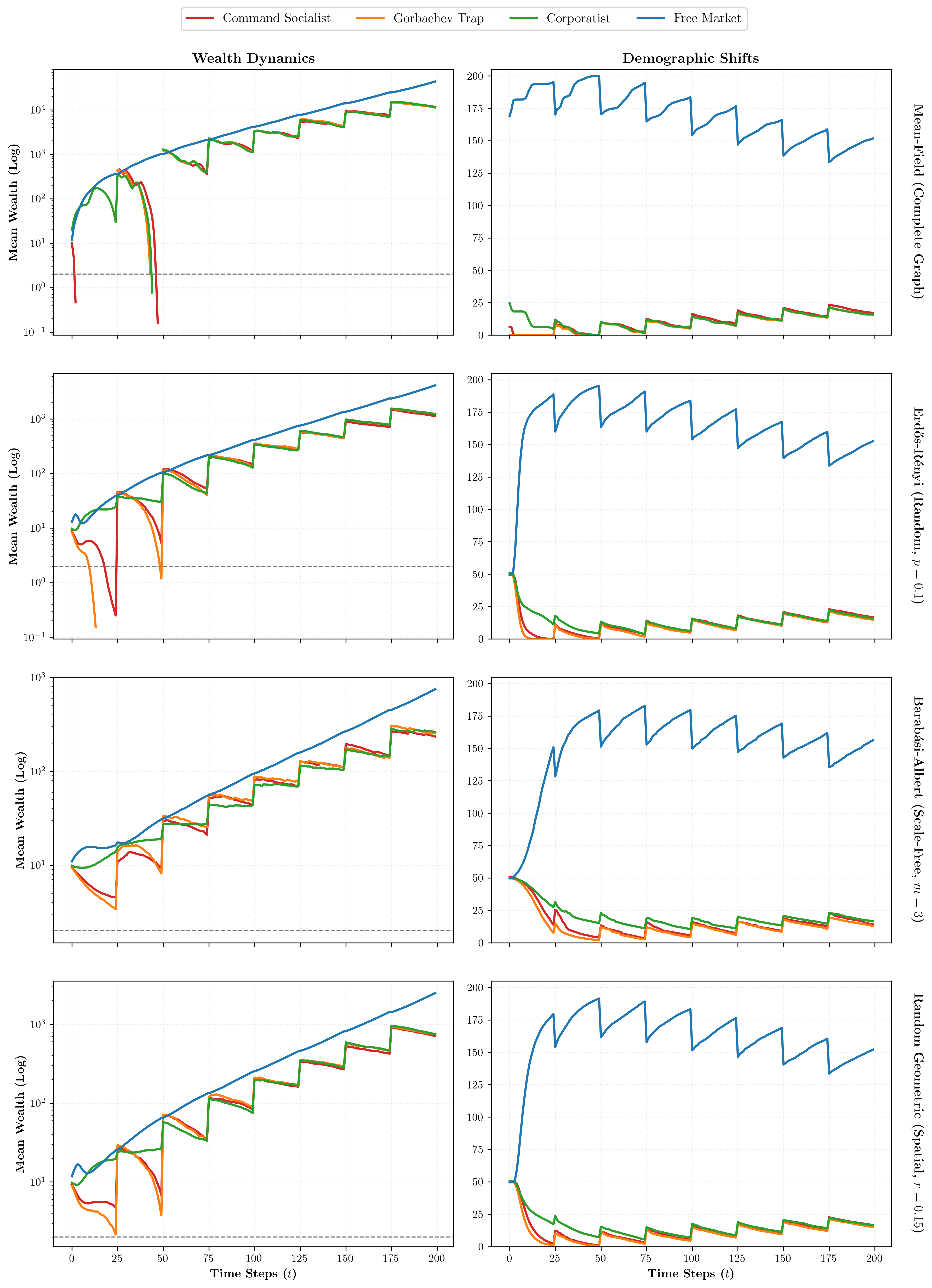}
    \caption{Comparative Monte Carlo ensemble ($N=200$, $t=300$ steps) demonstrating the macroscopic phase evolution of institutional factions across four distinct topological configurations. The top row illustrates mean wealth (logarithmic scale) bounded by $\pm 1\sigma$ standard deviation, while the bottom row visualizes demographic dominance.}
    \label{fig:topology}
\end{figure}

To evaluate the emergent properties of the SNG model, we executed Monte Carlo simulations over $R=100$ independent realizations for $t=300$ steps (for theoretical topologies) and $t=47$ steps (for the empirical historical replay). Figure \ref{fig:topology} overlays the mean wealth trajectories and demographic survival of the hegemonic faction across distinct network architectures.

The results reveal a profound divergence in systemic risk. In the democratized influence structures of the Mean-Field (Complete) and Erdős-Rényi graphs, structural shocks are rapidly diluted. When random nodes collapse, their inefficient capital is absorbed by the healthy majority, resulting in tight $\pm 1\sigma$ variance bands and predictable J-Curve recoveries. 

Conversely, the Barabási-Albert (Scale-Free) network introduces severe systemic fragility, which we define as the \textit{Hub-Risk Paradigm}. Because influence is centralized, the macroscopic trajectory is highly sensitive to the locus of the exogenous shock. If a stochastic revolution infects a high-degree ``mega-hub,'' ideological friction propagates non-linearly, artificially sustaining structurally defunct ``Zombie Factions'' and permanently suppressing the derivative of global TFP.

\subsection{Spatial Firewalls in Geometric Networks}

\begin{figure}[htbp]
    \centering
    \includegraphics[width=\textwidth]{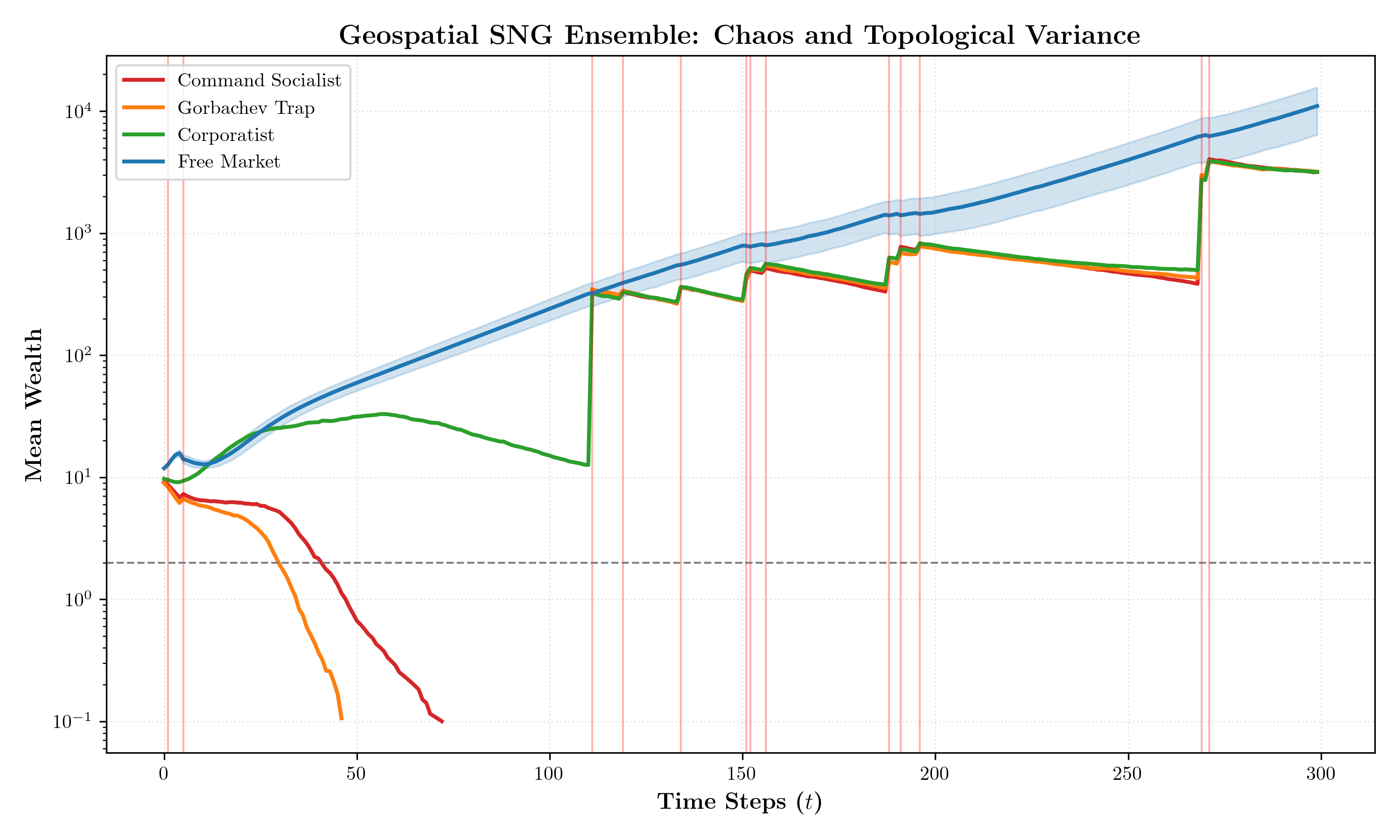}
    \caption{Phase evolution in the Random Geometric Graph. Physical boundaries heavily restrict rapid network integration, prolonging the survival of inefficient factions (Command Socialist/Corporatist) compared to mean-field topologies, acting as an epidemiological firewall against contagion.}
    \label{fig:geospatial}
\end{figure}

While scale-free networks act as super-spreaders during geopolitical volatility, physical geography provides systemic immunity. Figure \ref{fig:geospatial} isolates the mechanics of the Random Geometric Graph (RGG).

While the RGG's absolute compounding is slightly restricted by localized trade limitations, its variance remains exceptionally tight during crises. Physical spatial constraints act as a \textit{Spatial Firewall}. When a shock triggers an institutional collapse, the ideological contagion is physically quarantined. Unaffected regions continue to compound capital undisturbed until they possess sufficient wealth to systematically re-assimilate the collapsed border nodes, preventing systemic global contagion.

\subsection{Empirical Historical Replay (1970--2017)}

\begin{figure}[htbp]
    \centering
    \includegraphics[width=1.0\textwidth]{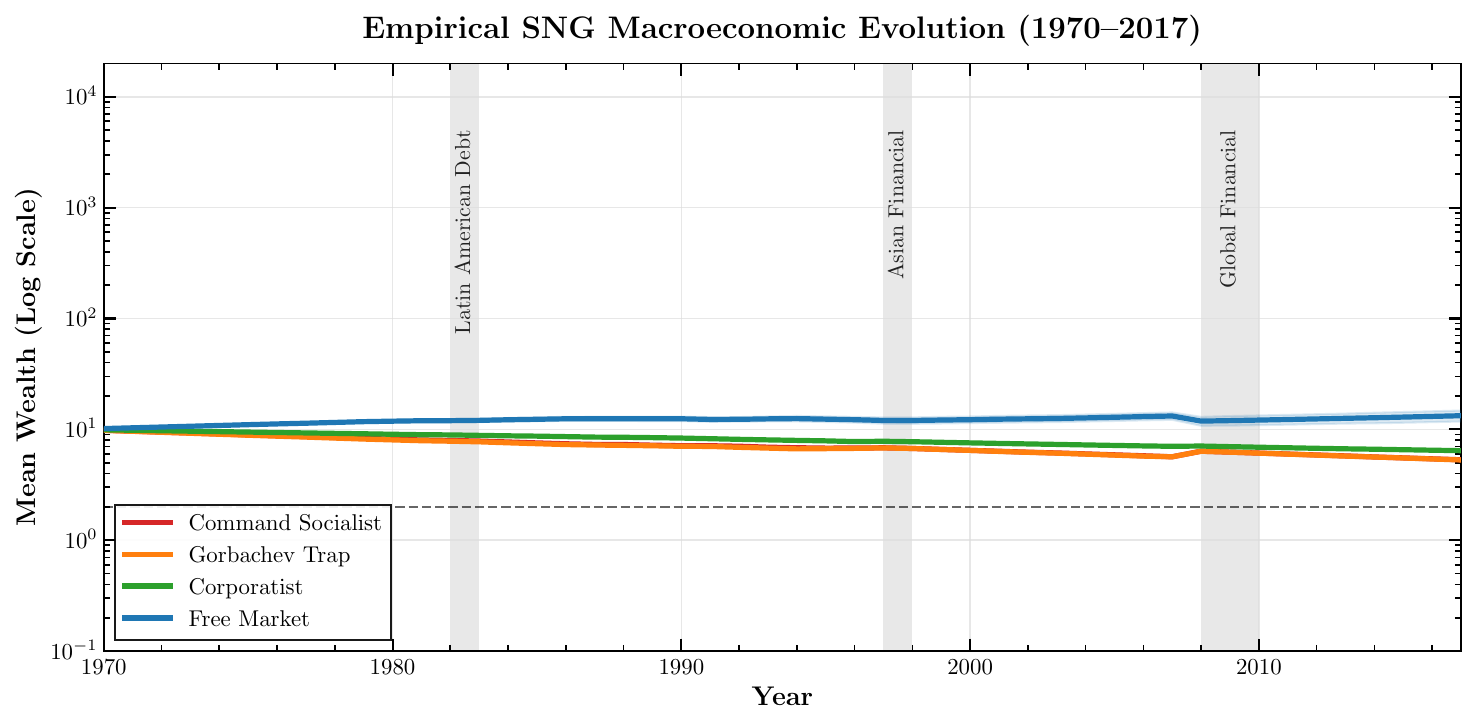}
    \caption{Empirical historical simulation (1970--2017) using CEPII Gravity trade topology ($N=100$) and Laeven-Valencia targeted exogenous shocks. Shaded regions denote major systemic crises. The model endogenously recreates the 1989--1991 Soviet collapse and demonstrates the expanding variance ($\pm 1\sigma$) following the 2008 Global Financial Crisis.}
    \label{fig:historical}
\end{figure}

To test the SNG model's predictive validity, we executed the simulation using the empirical CEPII Gravity topology and the IMF systemic crisis timeline. Figure \ref{fig:historical} displays the mean wealth evolution of competing institutional factions over this 47-year period.

The simulation accurately reproduces the catastrophic phase transition of the late 20th century without hardcoded chronological triggers. Factions seeded with severe institutional contradictions---specifically those mirroring the ``Gorbachev Trap'' $(0,1)$---experience compounding decay. Because their TFP falls below the depreciation rate $d$, they hemorrhage capital to more efficient neighbors and breach the panic threshold, successfully modeling the deterministic collapse of the Soviet economic bloc between 1989 and 1991 as a thermodynamic necessity.

A critical validation of the SNG engine is its endogenous handling of the ``China Exception.'' While the model accurately forced the catastrophic collapse of states trapped in rigid Command Socialist configurations $[0,0]$ (such as the USSR), it mathematically allowed for the survival of nations executing sequenced, partial reforms. Empirically, post-1978 China mutated away from a pure command economy into a Corporatist / State Capitalist architecture $[1,0]$. By liberalizing internal price mechanisms and trade while maintaining strict state-directed financial controls, this specific genomic configuration avoided the fatal ``Gorbachev Trap'' ($P_G$). Consequently, the SNG engine allowed the Corporatist faction to maintain sufficient Total Factor Productivity to outpace baseline depreciation ($d$). It survived the massive volatility of the 1990s without breaching the systemic $\tau_{\text{panic}}$ threshold, directly mirroring China's historical macroeconomic divergence from the Soviet bloc and validating the precision of the discrete genomic fitness landscape.

Furthermore, the integration of empirical IMF crisis dates exposes the real-world mechanics of the J-Curve. The vertical shaded regions mark the targeted systemic crises of 1982, 1997, and 2008. During these exact epochs, the Free Market wealth trajectory registers distinct structural dips, accompanied by massive expansions in the $\pm 1\sigma$ variance band. This expansion indicates permanent structural scarring within the topology. However, because the CEPII Gravity network acts as an empirical spatial firewall, the localized shocks are quarantined. The Free Market hegemon absorbs the localized damage, re-assimilates the collapsed nodes, and resumes its exponential compounding trend, mathematically demonstrating the extreme resilience of decentralized, positive-sum market networks.

\section{Conclusion}

In this paper, we introduced the Stochastic Networked Governance (SNG) model, a positive-sum, agent-based computational framework designed to capture the non-linear dynamics of macroeconomic regime change. By moving beyond the frictionless equilibrium assumptions of neoclassical growth models, the SNG engine successfully formalizes the mechanisms of institutional path dependency, endogenous value creation, and the thermodynamic friction of structural reform. 

Our Monte Carlo simulations demonstrate that when macroeconomic policy is modeled as a discrete genomic vector, the global economy exhibits deterministic phase transitions. Factions burdened by severe institutional complementarity failures suffer catastrophic TFP decay and are rapidly eliminated via spatial Tiebout sorting. In a low-volatility environment, the system reliably converges upon a unipolar hegemony, as capital flows strictly toward the network's most efficient institutional attractor, allowing for uninterrupted exponential compounding.

By replacing theoretical abstractions with the empirical CEPII Gravity network and the Laeven-Valencia crisis timeline, the SNG engine successfully functioned as a macroeconomic historical simulation. The model endogenously replicated the phase transition and collapse of Command Socialist regimes in the late 20th century. Furthermore, it proved that the topological constraints of distance and economic mass act as a "Spatial Firewall," quarantining localized contagion during systemic events like the 1997 Asian Financial Crisis and allowing the global network to recover its compounding trajectory.

While the SNG model provides a robust baseline for evaluating institutional evolution, it currently operates under the assumption of a homogeneous geographic landscape. Future research will expand this framework by introducing heterogeneous natural resource endowments to test the ``Resource Curse'' hypothesis, exploring whether exogenous geographic wealth can subsidize inefficient institutions and delay inevitable regime collapse. Additionally, conducting bivariate parameter sweeps of the ideological trade friction coefficient ($\gamma$) will allow us to model localized economic embargoes and the formation of polarized, multipolar spheres of influence. 

Ultimately, the SNG framework bridges econophysics and institutional economics, providing a rigorous mathematical foundation for understanding why human governance fails, how it recovers, and the topological limits of global economic resilience.

\section*{Declaration of Generative AI and AI-assisted Technologies}
During the preparation of this work, the author(s) used Gemini to program the simulation and assist in drafting the manuscript. After using this tool/service, the author(s) reviewed and edited the content as needed and take(s) full responsibility for the content of the published article.


\begin{thebibliography}{99}

\bibitem{solow1956contribution}
Solow, R. M. (1956). A contribution to the theory of economic growth. \textit{The Quarterly Journal of Economics}, 70(1), 65--94.

\bibitem{romer1990endogenous}
Romer, P. M. (1990). Endogenous technological change. \textit{Journal of Political Economy}, 98(5, Part 2), S71--S102.

\bibitem{north1990institutions}
North, D. C. (1990). \textit{Institutions, Institutional Change and Economic Performance}. Cambridge University Press.

\bibitem{acemoglu2005institutions}
Acemoglu, D., Johnson, S., \& Robinson, J. A. (2005). Institutions as a fundamental cause of long-run growth. \textit{Handbook of Economic Growth}, 1, 385--472.

\bibitem{hellman1998winners}
Hellman, J. S. (1998). Winners take all: the politics of partial reform in postcommunist transitions. \textit{World Politics}, 50(2), 203--234.

\bibitem{arthur1999complexity}
Arthur, W. B. (1999). Complexity and the economy. \textit{Science}, 284(5411), 107--109.

\bibitem{jackson2010social}
Jackson, M. O. (2010). \textit{Social and Economic Networks}. Princeton University Press.

\bibitem{epstein1996growing}
Epstein, J. M., \& Axtell, R. (1996). \textit{Growing Artificial Societies: Social Science from the Bottom Up}. Brookings Institution Press.

\bibitem{tesfatsion2002agent}
Tesfatsion, L. (2002). Agent-based computational economics: Growing economies from the bottom up. \textit{Artificial Life}, 8(1), 55--82.

\bibitem{delligatti2008emergent}
Delli Gatti, D., Gaffeo, E., Gallegati, M., Giulioni, G., \& Palestrini, A. (2008). \textit{Emergent Macroeconomics: An Agent-Based Approach to Business Fluctuations}. Springer Science \& Business Media.

\bibitem{mantegna1999introduction}
Mantegna, R. N., \& Stanley, H. E. (1999). \textit{Introduction to Econophysics: Correlations and Complexity in Finance}. Cambridge University Press.

\bibitem{bouchaud2000wealth}
Bouchaud, J. P., \& Mézard, M. (2000). Wealth condensation in a simple model of economy. \textit{Physica A: Statistical Mechanics and its Applications}, 282(3-4), 536--545.

\bibitem{yakovenko2009colloquium}
Yakovenko, V. M., \& Rosser, J. B. (2009). Colloquium: Statistical mechanics of money, wealth, and income. \textit{Reviews of Modern Physics}, 81(4), 1703.

\bibitem{castellano2009statistical}
Castellano, C., Fortunato, S., \& Loreto, V. (2009). Statistical physics of social dynamics. \textit{Reviews of Modern Physics}, 81(2), 591.

\bibitem{kauffman1993origins}
Kauffman, S. A. (1993). \textit{The Origins of Order: Self-Organization and Selection in Evolution}. Oxford University Press.

\bibitem{tiebout1956pure}
Tiebout, C. M. (1956). A pure theory of local expenditures. \textit{Journal of Political Economy}, 64(5), 416--424.

\bibitem{head2014gravity}
Head, K., \& Mayer, T. (2014). Gravity equations: Workhorse, toolkit, and cookbook. \textit{Handbook of International Economics}, 4, 131--195.

\bibitem{conte2022cepii}
Conte, M., Cotterlaz, P., \& Mayer, T. (2022). \textit{The CEPII Gravity Database}. CEPII Working Paper N°2022-05, July 2022. Available at: \url{http://www.cepii.fr/CEPII/en/bdd_modele/bdd_modele_item.asp?id=8}.

\bibitem{barabasi1999emergence}
Barabási, A. L., \& Albert, R. (1999). Emergence of scaling in random networks. \textit{Science}, 286(5439), 509--512.

\bibitem{laeven2018systemic}
Laeven, M. L., \& Valencia, M. F. (2018). Systemic banking crises revisited. \textit{International Monetary Fund Working Papers}, WP/18/206.

\end{thebibliography}
\end{document}